\documentclass[12pt]{iopart}
\usepackage{physics}
\usepackage{color}
\usepackage{here}
\usepackage{graphicx}
\usepackage{amsmath,amssymb,mathtools,bm}

\begin{document}
\title{von Neumann entropy of phase space structures in gyrokinetic plasma turbulence}

\author{Go Yatomi$^1$, Motoki Nakata$^{2,3}$}
\ead{yatomi.go@nifs.ac.jp}
\address{$^1$ National Institutes of Natural Sciences, National Institute for Fusion Science, Toki, Gifu 509-5292, Japan}
\address{$^2$ Faculty of Arts and Sciences, Komazawa University, Tokyo 154-8525, Japan}
\address{$^3$ RIKEN Center for Interdisciplinary Theoretical and Mathematical Sciences (iTHEMS), Saitama 351-0198, Japan}

\begin{abstract}
  We introduce a data--driven diagnostic that combines the singular value decomposition (SVD) with an information-theoretic entropy to quantify the phase-space complexity of perturbed distribution functions in gyrokinetic turbulence. Applying this framework to nonlinear flux-tube simulations that solve the time evolution of the ion distribution function represented by Fourier modes with the wavenumber for real space, we define the von Neumann entropy (vNE) to analyze the velocity-space structure. A global survey in the wavenumber space reveals a wavenumber-dependent variation of the vNE in velocity-space structure: the vNE remains low at low wavenumber but increases across \(k_\perp\rho_{t\mathrm{i}}\sim 1\). Hermite/Laguerre decompositions revealed that the finite Larmor radius (FLR) phase mixing in the perpendicular (magnetic-moment) direction is active. Simultaneously, the systematic increase in vNE for \(k_\perp\rho_{t\mathrm{i}}\) correlates with the broadening of the Hermite spectrum, suggesting enhanced parallel phase mixing (Landau resonance) as the primary mechanism for the observed wave number dependence. These results demonstrate that the SVD-based vNE provides a compact measure of kinetic complexity without assuming a predefined basis and enables a global mapping of its wavenumber dependence of phase-mixing processes in gyrokinetic turbulence.
\end{abstract}

\maketitle
\section{Introduction} \label{sec:introduction}
To understand the physics of turbulence, which possesses complex properties such as multiscale behavior, inhomogeneity, and nonlinearity, it is necessary to develop methods for extracting essential information from data and translating it into interpretable physical quantities. Modal decompositions play a central role in this endeavor—most notably proper orthogonal decomposition (POD) and related techniques—by isolating energetically dominant structures and providing low-order coordinates for analysis and modeling \cite{Holmes1996,Schmid2010,Taira2017}. Such tools, including POD, singular value decomposition (SVD), and dynamic mode decomposition (DMD), have been successfully employed across plasma turbulence \cite{Sasaki2019,Sasaki2021,Maeyama2021a,Yatomi2023} and classical fluid dynamics \cite{Lim2004,Ravindran2000,Pan2011}. In parallel with energetics-based analyses, information-theoretic approaches have recently been explored in classical many-body and fluid turbulence \cite{Falkovich2023,Tanogami2024,Pizzi2024}. In quantum systems, closely related concepts, i.e., von Neumann/Shannon entropies and Schmidt decompositions (i.e., SVD), are foundational, and research using information entropy continues in various physical systems such as black holes and spin many-body systems \cite{Horodecki2009,Belokolos2009,Matsueda2012}. Motivated by the mathematical similarity between data-driven mode decompositions and quantum states, in our previous work, we have defined information entropies of turbulent fields based on the SVD and shown that it captures nontrivial state transitions and effectively reduces the dimension or information of nonlinear interactions in plasma-fluid systems \cite{Yatomi2025}. \par

A second key thread concerns the physics of \emph{phase-space turbulence} in gyrokinetic plasmas, where free energy is exchanged not only across real-spatial scales but also across velocity-space scales by linear and nonlinear phase mixing processes. Foundational theory predicted cascades of gyrokinetic free energy (or Gibbs entropy) mediated by perpendicular nonlinear interactions and by parallel Landau resonances, with characteristic signatures in Hermite spectra and in the distribution of energy among velocity moments \cite{Tatsuno2009,Plunk2010,Schekochihin2016,Mandell2018}. Furthermore, coherent phase-space structures, which have long been known from the Bernstein-Greene-Kruskal (BGK) construction, have re-emerged as organizing elements of drift-wave turbulence, intermittency, and transport modeling \cite{Bernstein1957,Lesur2014,Kosuga2013a,Kosuga2017}. These advances indicate that any faithful measure of the complexity of kinetic turbulence must resolve not only how fluctuations distribute across perpendicular wavenumber, which sets the cascade pathway in real space, but also how they populate velocity space where Landau resonances and finite Larmor radius (FLR) phase mixing generate finer structure. Yet, because distribution functions are inherently high-dimensional and their volume of data is enormous, systematic and global investigation in the wavenumber space of the predicted velocity-space complexity has been scarce. \par

In this paper, we address this gap by analyzing the vNE, which does not assume \emph{a priori} basis choices and, instead, uses the characteristic singular vectors of the dataset as the natural Hilbert basis. This strategy lets us map, across the entire wavenumber space, where only a few modes suffice and where many are required, thereby revealing a previously unresolved, scale-dependent transition of velocity-space complexity. Moreover, we place our approach in relation to conventional velocity-space diagnostics. Hermite and Laguerre projections diagnose, respectively, parallel (Landau) and perpendicular (FLR) phase mixing in the perpendicular (magnetic-moment) direction, and their spectral slopes and fluxes quantify the direction and strength of inter-scale transfers \cite{Schekochihin2016,Mandell2018}. In this work we use those projections alongside vNE to disentangle which mechanisms control the observed rise of complexity at high wavenumber and to connect the low-wavenumber regime to drift-wave instabilities identified by linear gyrokinetic growth rates. \par

The remainder of this paper is organized as follows. In Sec.~\ref{sec:vNE}, the algorithm of the SVD and the definition of the vNE are explained. The gyrokinetic-Poisson equation system and its numerical simulation are described in Sec.~\ref{sec:GKE}. Then, the SVD is applyed to the distribution function in the velocity space, and the result of the dependence of the vNE of the distribution function on the wavenumber space is presented in Sec.~\ref{sec:result}. In Sec.~\ref{sec:discussion}, we discuss the result of the vNE of the distribution function, comparing the Hermite/Laguerre decomposition. Finally, the concluding remarks are given in Sec.~\ref{sec:summary}.

\section{Theory of modal analysis and numerical setup} \label{sec:method}
\subsection{Singular value decomposition (SVD) for time-evolving 2D fields and construction of von Neumann entropy (vNE)} \label{sec:vNE}
As a compact diagnostic for high-dimensional kinetic fields, we define the von
Neumann entropy (vNE) of turbulent fields by combining singular value decomposition
(SVD) with information-theoretic weighting of modes \cite{Yatomi2025}. \par
Let $f(x,y,t)$ denote a two-dimensional field sampled on a uniform grid
$\{(x_i,y_j)\}_{i=1,\dots,N_x;\,j=1,\dots,N_y}$ and at $N_t$ time frames $\{t_n\}_{n=1}^{N_t}$.
Each snapshot of the field is reshaped to a vector
\[
\vb*{f}(t_n)\in\mathbb{C}^{N_xN_y},\qquad
\vb*{f}(t_n)=\bigl(f(x_1,y_1,t_n),\cdots,f(x_{N_x},y_{N_y},t_n)\bigr)^{\mathsf T}.
\]
Stacking all fields and times column-wise yields the data matrix
\[
F\in\mathbb{C}^{N_xN_y\times N_t},\qquad
F=\bigl(\vb*{f}(t_1),\cdots,\vb*{f}(t_{N_t})\bigr).
\]
Applying SVD,
\[
F=\,U\,\Sigma\,V^\dagger,\qquad
\Sigma=\mathrm{diag}(s_1,\cdots,s_N),
\]
where $N=\min\{N_xN_y,\,N_t\}$, yields orthonormal spatial modes (columns of $U$) and orthonormal temporal coefficients (columns of $V$). The columns of $U$ are the eigenvectors of the spatial covariance
$\mathbf{F}\mathbf{F}^\dagger$, and the columns of $V$ are the eigenvectors of the temporal
covariance $\mathbf{F}^\dagger\mathbf{F}$, with $\Sigma^2$ containing the corresponding
eigenvalues. Writing $\psi_i(x_j,y_k)=U_{j+N_y(k-1),i}$ and $h_i(t_n)=V_{n,i}$, the field expands as
\begin{equation}
  f(x,y,t)=\sum_{i=1}^{N} s_i\,h_i(t)\,\psi_i(x,y),
  \label{eq:svd-expansion}
\end{equation}
where the diagonal component $s_i$ of the matrix $\Sigma$ are referred to as the singular value. \par

Because $\{\psi_i\}$ are orthonormal, i.e.,
\begin{equation}
  \int \psi_i(x,y)\,\psi_j(x,y)\,\dd^2\vb*{x}=\delta_{ij}, \label{eq:orthonormal}
\end{equation}
they can be regarded as a unitary basis in an
$(N_xN_y)$-dimensional Hilbert space of physical fields. Selecting a subset $X\subset\{1,\dots,N\}$ (e.g.,
a mode window, a $k$-shell, or any mask in the SVD index), we define a mixed state as a density matrix through the mode weights $\{\eta_i\}_{i\in X}$:
\begin{equation}
  \rho_X \;=\; \sum_{i\in X} \eta_i\,\dyad{\psi_i}{\psi_i}, \label{eq:density-matrix}
\end{equation}
where
\begin{equation}
  \eta_i \;=\;
  \frac{\displaystyle \bigl|s_i\,h_i(t)\bigr|^2}
       {\displaystyle \sum_{j\in X}\bigl|s_j\,h_j(t)\bigr|^2},
  \quad \sum_{i\in X}\eta_i=1.
  \label{eq:eta_svd}
\end{equation}
The von Neumann entropy (vNE) associated with $X$ is then
\begin{equation}
  S^{(X)}_{\mathrm{vN}} \;=\; -\Tr\,\bigl(\rho_X\ln\rho_X\bigr)
  \;=\; -\sum_{i\in X}\eta_i\ln\eta_i.
  \label{eq:vne}
\end{equation}
Because $h_i(t)$ are the components of the normalized right singular vectors
(i.e.\ columns of the unitary matrix $V$), the weights $\eta_i$ reduce to normalized fractions over the singular values; hence, $S_{\mathrm{vN}}^{(X)}$ is a Shannon-type entropy of the SVD spectrum for the chosen subset $X$. \par
The von Neumann entropy $S_{\mathrm{vN}}^{(X)}$ in \eqref{eq:vne} quantifies how many SVD modes are effectively required to represent the field within the chosen subset $X$.  If the energy or the contribution fraction to the original field concentrates on a few singular modes, the weights $\{\eta_i\}_{i\in X}$ are strongly peaked and $S_{\mathrm{vN}}^{(X)}$ is small, indicating a low-rank, coherent structure.  Conversely, when the energy spreads over many modes, $S_{\mathrm{vN}}^{(X)}$ increases, signaling a high-rank, strongly mixed state. Importantly, the basis functions $\{\psi_i\}$ are obtained by data-driven analysis: they are learned from the snapshots via SVD as orthonormal eigenmodes of the sample covariance ($U$ are the eigenvectors of $FF^\dagger$ and $V$ are those of $F^\dagger F$), not prescribed \emph{a priori} like Fourier harmonics or a probability-density basis.  Hence $S_{\mathrm{vN}}^{(X)}$ measures the intrinsic complexity of the observed dynamics in the Hilbert space spanned by the data themselves, avoiding bias from a fixed representation.  In this sense, vNE plays the role of a Shannon-type entropy on the SVD spectrum: it counts how broadly the normalized right singular-vector components $h_i(t)$ distribute energy across singular values and thereby how rich a set of spatio-temporal patterns is needed to reconstruct the field within $X$.

\subsection{Gyrokinetic equation and its numerical simulation} \label{sec:GKE}
We employ the electrostatic, kinetic descriptions of magnetized plasma turbulence based on the nonlinear gyrokinetic equation for the perturbed gyrocenter distribution function $\delta f_{\mathrm{s}\bm k_\perp}(z,v_\parallel,\mu,t)$ of species 's', coupled to the Poisson equation for the electrostatic potential $\delta\phi_{\bm k_\perp}(t)$. For numerical solution we use the Eulerian gyrokinetic Vlasov code GKV, originally developed for flux-tube simulations of drift-wave turbulence in toroidal plasmas~\cite{Watanabe2006}. \textsc{GKV} solves the gyrokinetic Vlasov equation in $(k_x,k_y,z,v_\parallel,\mu)$ phase space with spectral and finite-difference discretizations, model (or linearized Landau) collisions, and de-aliasing for the nonlinear convolution. The gyrokinetic-Poisson equations in Fourier wavenumber representation are described as
\begin{align}
\Biggl[
\frac{\partial}{\partial t}
&+ v_\parallel \nabla_\parallel
+ i\omega_{\mathrm{Ds}} 
- \qty(\frac{e_{\mathrm{s}} \mu}{m_i}\nabla_\parallel B)\frac{\partial}{\partial v_\parallel}
\Biggr]\delta f_{\mathrm{s}\bm k_\perp} \nonumber \\
&
-\frac{1}{B}
\sum_{\bm k'_\perp+\bm k''_\perp=\bm k_\perp}
\bigl[\bm b\cdot (\bm k'_\perp \times \bm k''_\perp)\bigr]
\,J_0(k'_\perp\rho_{t\mathrm{s}})\,\delta\phi_{\bm k'_\perp}\,\delta f_{\mathrm{s}\bm k''_\perp}
\nonumber\\
&=
\frac{e_{\mathrm{s}} F_{M\mathrm{s}}}{T_{\mathrm{s}}}
\Bigl(i\omega_{\ast T\mathrm{s}}+i\omega_{\mathrm{Ds}}-v_\parallel\nabla_\parallel\Bigr)
J_0(k_\perp\rho_{t\mathrm{s}})\,\delta\phi_{\bm k_\perp}
\;+\; C_{\mathrm{s}},
\label{eq:gke}
\end{align}
\begin{align}
\qty[
k_\perp^2
+\frac{1}{\varepsilon_0}\sum_{\mathrm{s}} \frac{e_{\mathrm{s}}^2 n_{\mathrm{s}}}{T_{\mathrm{s}}}\qty(1-\Gamma_{0\mathrm{s}\,k_\perp})
]\delta\phi_{\bm k_\perp}
&=
\frac{1}{\varepsilon_0}\sum_{\mathrm{s}} e_{\mathrm{s}}
\int\! d^3v \, J_0(k_\perp\rho_{t\mathrm{s}})\,\delta f_{\mathrm{s}\bm k_\perp}.
\label{eq:qn}
\end{align}
Here, $\vb*{k}_\perp=(k_x,k_y)$ is the perpendicular wavevector, $\vb*{b}$ is the unit
vector along the equilibrium magnetic field line, and $B$ is the magnetic-field
strength. The symbol $\delta\phi_{\vb*{k}_\perp}$ denotes the electrostatic potential
fluctuation at wavevector $\vb*{k}_\perp$. The magnetic moment is
$\mu = m_{\mathrm{s}} v_\perp^2/(2B)$, with $e_{\mathrm{s}}$ the species charge,
$m_{\mathrm{s}}$ the particle mass, and $T_{\mathrm{s}}$ the (equilibrium) temperature
of species $\mathrm{s}$. The drift frequency $\omega_{\mathrm{Ds}}$ represents curvature
and $\nabla B$ drifts along the field line, while $\omega_{\ast T\mathrm{s}}$ is the
diamagnetic frequency including the temperature-gradient drive. The species Larmor
radius is $\rho_{t\mathrm{s}} = v_{t\mathrm{s}}/\Omega_{\mathrm{s}}$ with thermal speed
$v_{t\mathrm{s}}=\sqrt{T_{\mathrm{s}}/m_{\mathrm{s}}}$ and gyrofrequency
$\Omega_{\mathrm{s}}=e_{\mathrm{s}} B/m_{\mathrm{s}}$. Finite Larmor radius (FLR) effects appear via the
zeroth-order Bessel function $J_0$ and the FLR response $\Gamma_0(b)=e^{-b} I_0(b)$,
where $I_0$ is the modified Bessel function and $b=(k_\perp\rho_i)^2$. The Maxwellian
background distribution is denoted by $F_M$, and $C_{\mathrm{s}}$ represents the model
collision operator. The Poisson equation (\ref{eq:qn}) balances the linear polarisation
response, $\sum_{\mathrm{s}} \tfrac{e_{\mathrm{s}}^{\,2} n_{\mathrm{s}}}{\epsilon_0 T_{\mathrm{s}}}
\,(1-\Gamma_{0\mathrm{s}})\,\delta\phi_{\vb*{k}_\perp}$, against the gyroaveraged charge density
$\sum_{\mathrm{s}} e_{\mathrm{s}} \int \! \mathrm{d}^3 v\, J_0(k_\perp\rho_{t\mathrm{s}})\,
\delta f_{\mathrm{s}\vb*{k}_\perp}$ (see \cite{Nakata2012,Nakata2016} for more details). Hereafter, the perpendibular wavenumber $\vb*{k}_\perp$ is represented by the dimensionless wavenumber $\vb*{k}=\vb*{k}_\perp\rho_{t\mathrm{i}}$ for simplicity. \par 

\begin{figure}[H]
  \centering
  \includegraphics[width=10cm]{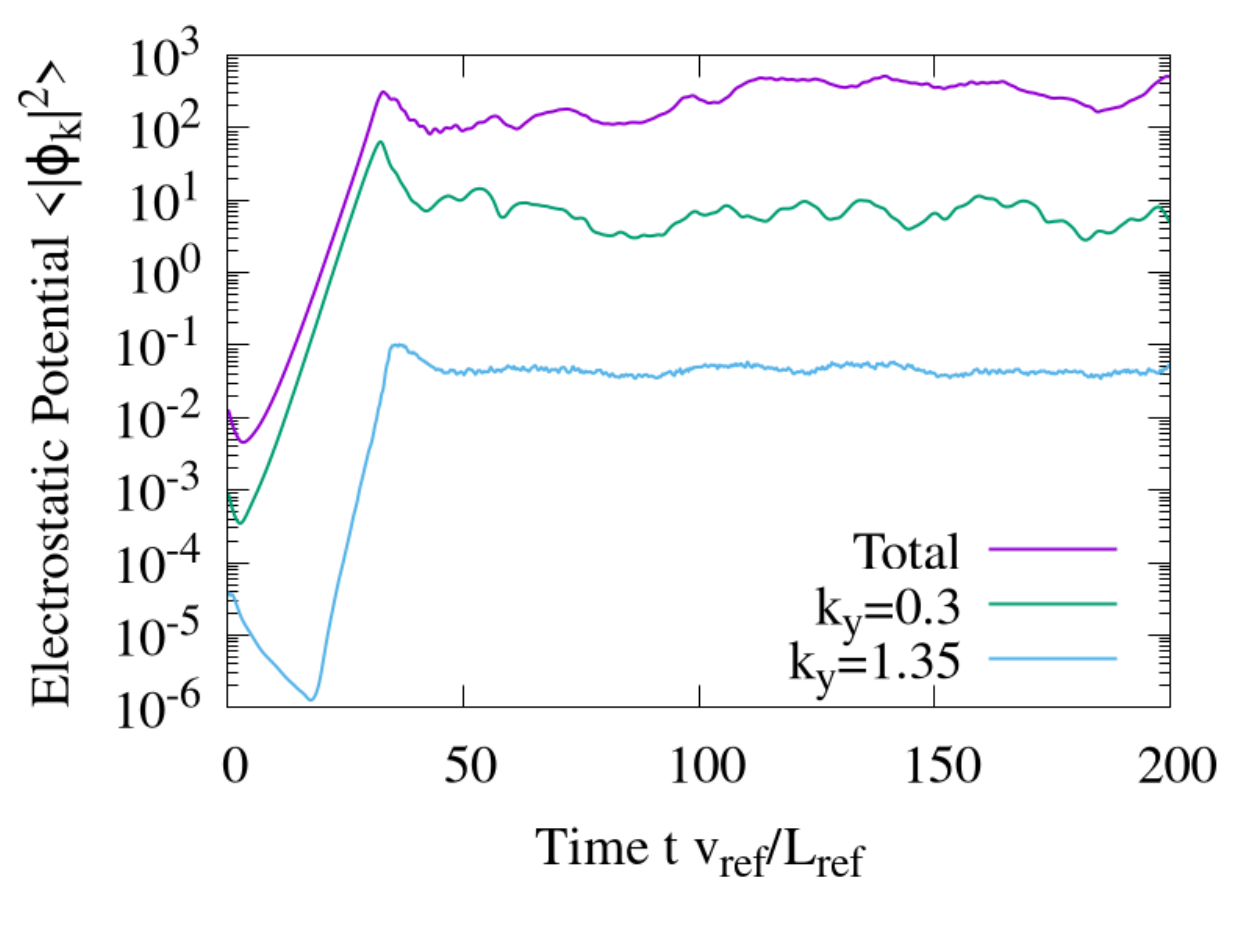}
  \caption{The temporal evolution of the electrostatic potential amplitude for the total value (purple), at $k_y=0.3$ (green), and $k_y=1.35$ (cyan).} 
  \label{fig:t-E}
\end{figure}

\begin{figure}[H]
  \centering
  \includegraphics[width=16cm]{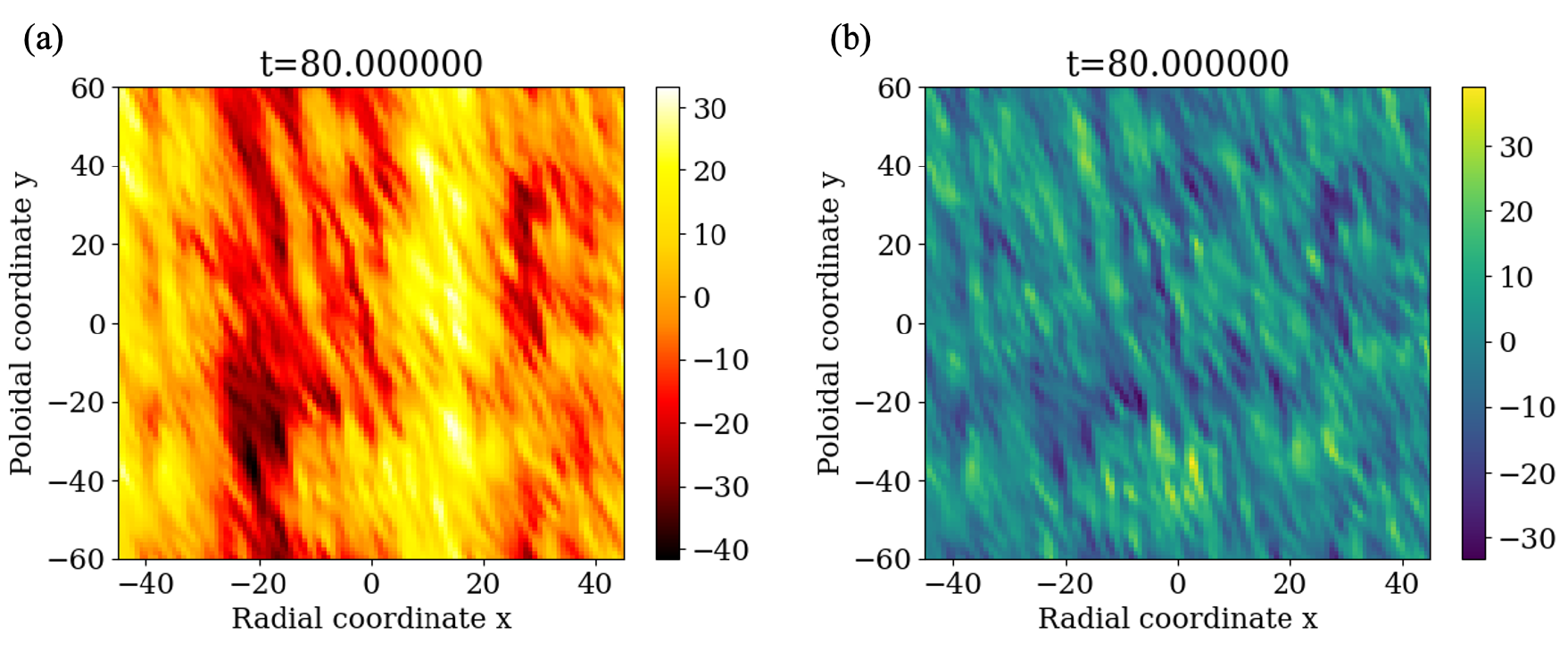}
  \caption{The snapshot of the turbulent fields in real space. (a) the electrostatic potential $\phi(x,y)$ and (b) the density fluctuation $n(x,y)$ at the time $t=80\,v_\mathrm{ref}/L_\mathrm{ref}$ and $z=0$.} 
  \label{fig:phi-n_contour}
\end{figure}

\begin{figure}[H]
  \centering
  \includegraphics[width=16cm]{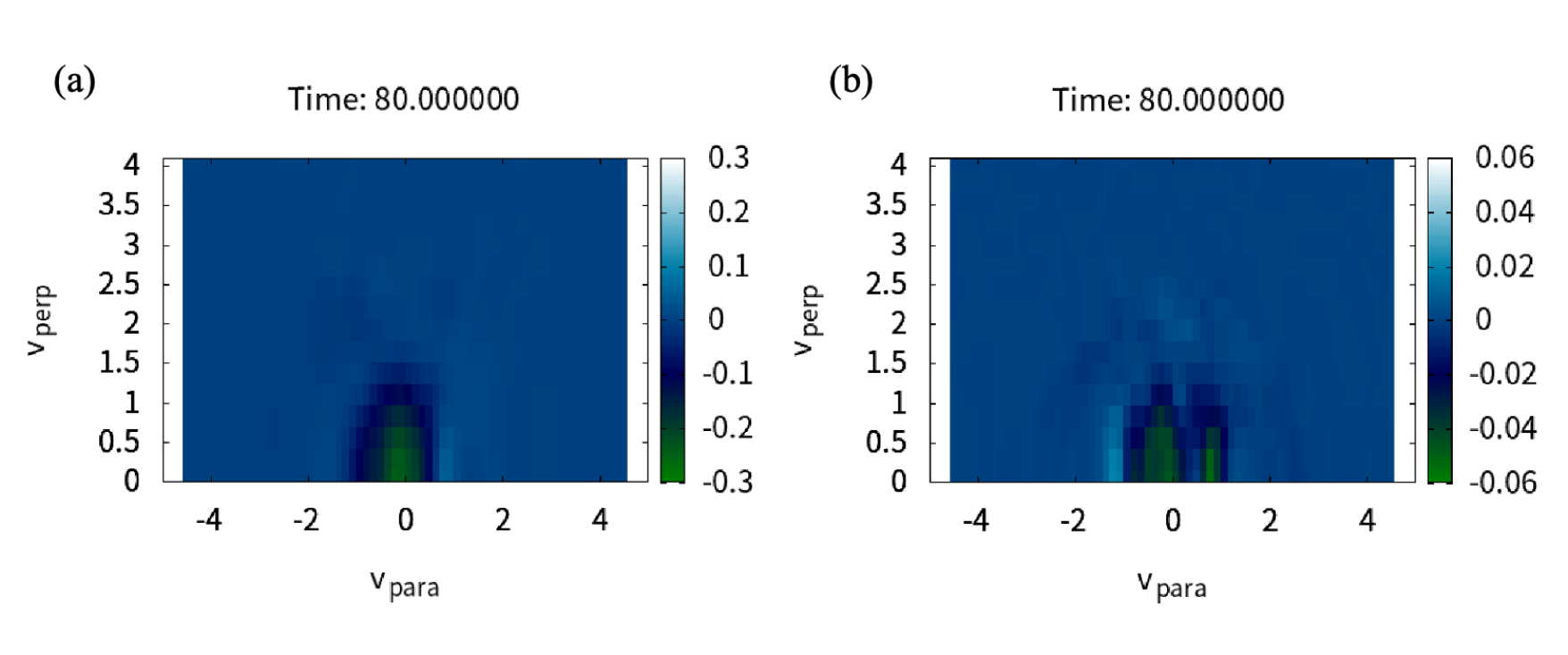}
  \caption{The snapshot of the distribution function $\mathrm{Re}[\delta f_{\mathrm{i}\bm k}(z,v_\parallel,\mu,t)]$. (a) the low-wavenumber case $(k_x,k_y)=(0,0.3)$ and (b) the high-wavenumber case $(k_x,k_y)=(0,1.35)$ at the time $t=80\,v_\mathrm{ref}/L_\mathrm{ref}$ and $z=0$.} 
  \label{fig:ff_contour}
\end{figure}

The simulations analyzed in this study adopt CBC-like core-tokamak parameters \cite{Dimits2000} (here, safety factor $q=1.4$, magnetic shear $\hat{s}=0.8$, and gradient drives $R/L_{T\mathrm{i}}=6.92$, $R/L_n=2.22$ in the conventional range for ITG turbulence), discretized on uniform $(k_x,k_y)$ grids and equidistant field-line coordinate $z$ with adequate resolution in $(v_\parallel,\mu)$. The nonlinear runs proceed from the linear-growth stage to a statistically stationary saturated state, as shown in Fig.~\ref{fig:t-E} which displays the temporal evolution of the electrostatic potential amplitude $\expval{|\delta\phi_{\vb*{k}}|^2}$ in total (purple), at $k_y=0.3$ (green), and $k_y=1.35$ (cyan). Figure~\ref{fig:phi-n_contour} presents contour plots of $\phi(x,y)$ and density fluctuations $n(x,y)=\int\! d^3v \, J_0(k)\,\delta f_{\mathrm{i}\bm k}$ at the time $t=80\,v_\mathrm{ref}/L_\mathrm{ref}$ and $z=0$. Here, the time is normalized by the reference length $L_\mathrm{ref}$ ($=R_a$ major radius at the magnetic axis) and the reference velocity $v_\mathrm{ref}=\sqrt{T_{\mathrm{i}}/m_{\mathrm{i}}}$. The velocity-space structure of the real copmonent of the distribution function $\mathrm{Re}[\delta f_{\mathrm{i}\bm k}(z,v_\parallel,v_\perp,t)]$ at $(k_x,k_y)=(0,0.3)$ and $(0,1.35)$ are illustrated in Fig.~\ref{fig:ff_contour}, highlighting that, in the saturated turbulence regime, appreciable fluctuations persist and their characteristic scales become finer toward higher wavenumbers, yielding increasingly sharp structure in velocity space.

\section{von Neumann entropy of the distribution function} \label{sec:result}
We analyze the nonlinear saturated phase by restricting the time window to $t\in[80,200]$ (see Fig.~\ref{fig:t-E}). For each wavevector $(k_x,k_y)$ we assemble the real part of the distribution function into a space--time data matrix and perform an SVD, such as
\begin{equation}
  \mathrm{Re}[\delta f_{\mathrm{i}\bm k}(v_\parallel,v_\perp,t)]=\sum_{i=1}^n\!s_i\,h_i(t)\,\psi_i(v_\parallel,v_\perp).
\end{equation}
Hereafter, we omit $\mathrm{Re}[\cdot]$ symbol for simplicity. Because the input is the fluctuating component, the velocity-space mean of the data is (by construction) zero at each $(k_x,k_y)$.

\begin{figure}[H]
  \centering
  \includegraphics[width=10cm]{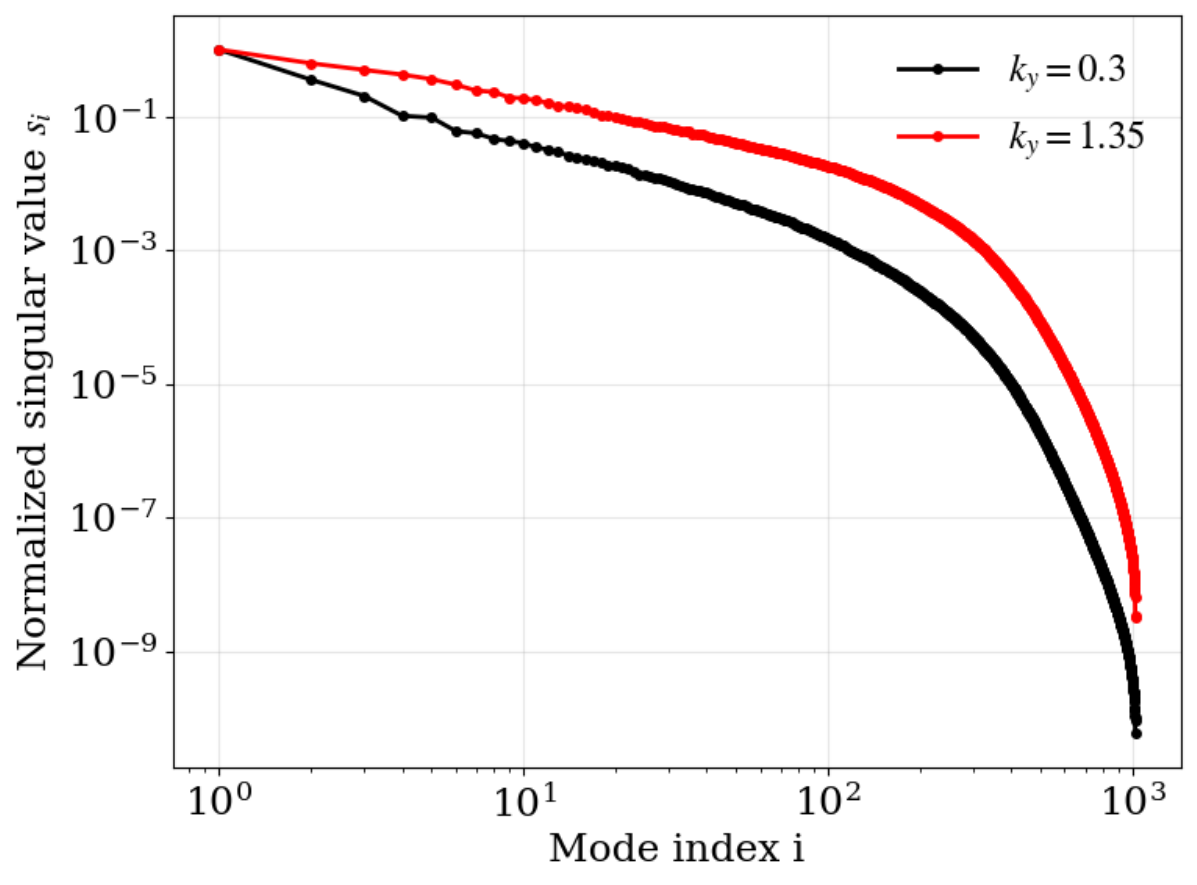}
  \caption{The spectra of the singular value $s_i$ in the SVD of the gyrokinetic distribution function $\delta f_{\mathrm{i}\bm k}(z,v_\parallel,\mu,t)$ at $(k_x,k_y)=(0,0.3)$ (black) and $(k_x,k_y)=(0,1.35)$ (red) normalized so that $s_1=1$.} 
  \label{fig:id-sv}
\end{figure}

Figure~\ref{fig:id-sv} shows representative singular-value spectra at a low-$k$ point $(k_x,k_y)=(0,0.3)$ and a high-$k$ point $(0,1.35)$ normalized so that $s_1=1$. Hereafter, the mode index is aligned in the descending order of the singular value $s_i$. The high-$k$ spectrum decays more slowly as the mode index increases, indicating broader modal support (i.e.\ a higher effective dimensionality) than at low $k$.

\begin{figure}[H]
  \centering
  \includegraphics[width=16cm]{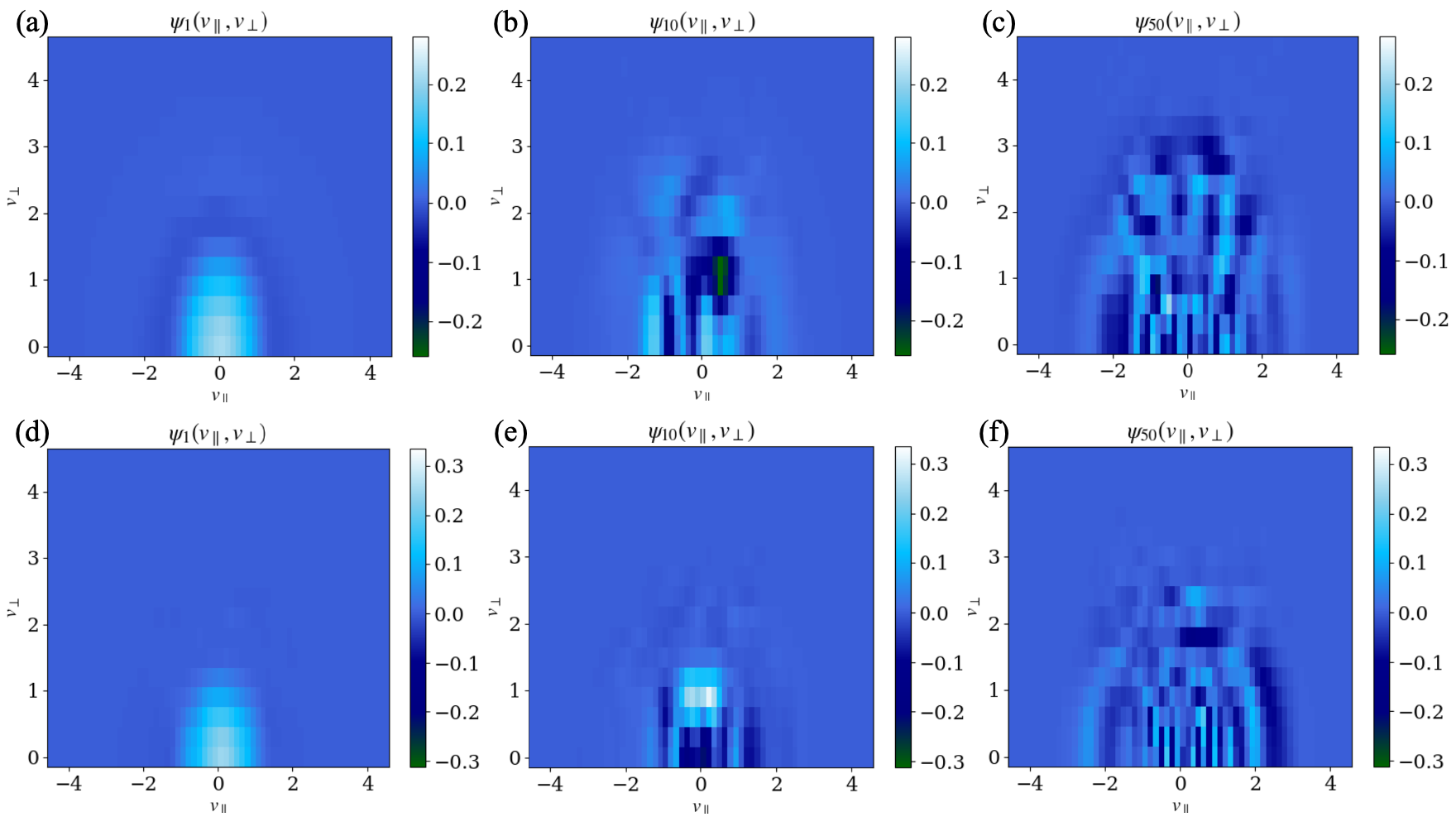}
  \caption{The velocity-space structure of the basis functions $\psi_i(v_\parallel,v_\perp)$ in the SVD of the gyrokinetic distribution function. } 
  \label{fig:vl-vp-psi}
\end{figure}

To visualize the associated velocity-space structures, Fig.~\ref{fig:vl-vp-psi} plots three leading spatial modes $\psi_i(v_\parallel,v_\perp)$ for the low-$k$ and high-$k$ cases. Clear fluctuating structures are observed; the characteristic scales in $(v_\parallel,v_\perp)$ become increasingly fine at larger mode index $i$, consistent with enhanced phase-space structuring in the saturated regime.

\begin{figure}[H]
  \centering
  \includegraphics[width=12cm]{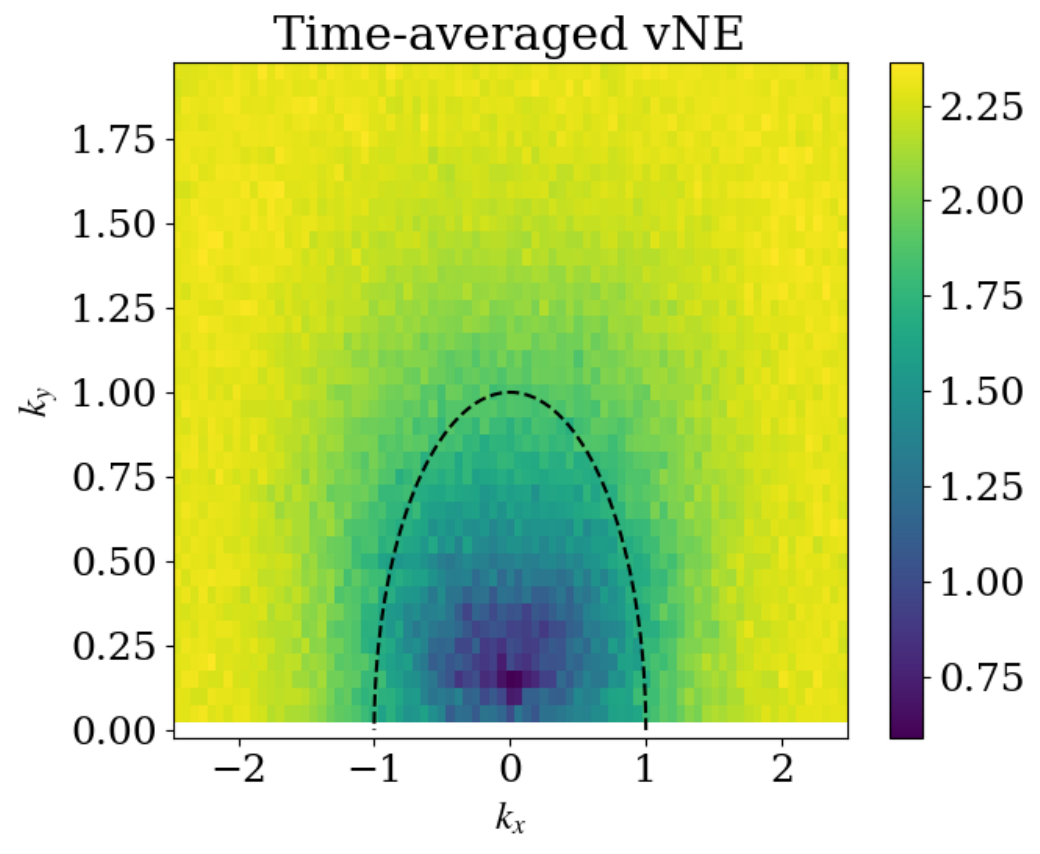}
  \caption{The contour in the wavenumber space of the von Neumann entropy (vNE) of the gyrokinetic distribution function. A semicircle for $k_x^2+k_y^2=1$ corresponding to the boundary of the small and large vNE region is shown as a black dashed curve.} 
  \label{fig:kx-ky-vNE}
\end{figure}

Using the SVD at each $(k_x,k_y)$, we compute the von Neumann entropy (vNE) from the time-dependent normalized spectral weights following Eq.~(\ref{eq:vne}) and then take a temporal average. The resulting contour map over the $(k_x,k_y)$ plane is shown in Fig.~\ref{fig:kx-ky-vNE}. The vNE is small at low wavenumber and large at high wavenumber, with a relatively sharp rise around $|\vb*{k}|\!\sim\!1$. In other words, low-$k$ fields are well represented by a few modes, whereas high-$k$ fields require many modes.

\begin{figure}[H]
  \centering
  \includegraphics[width=12cm]{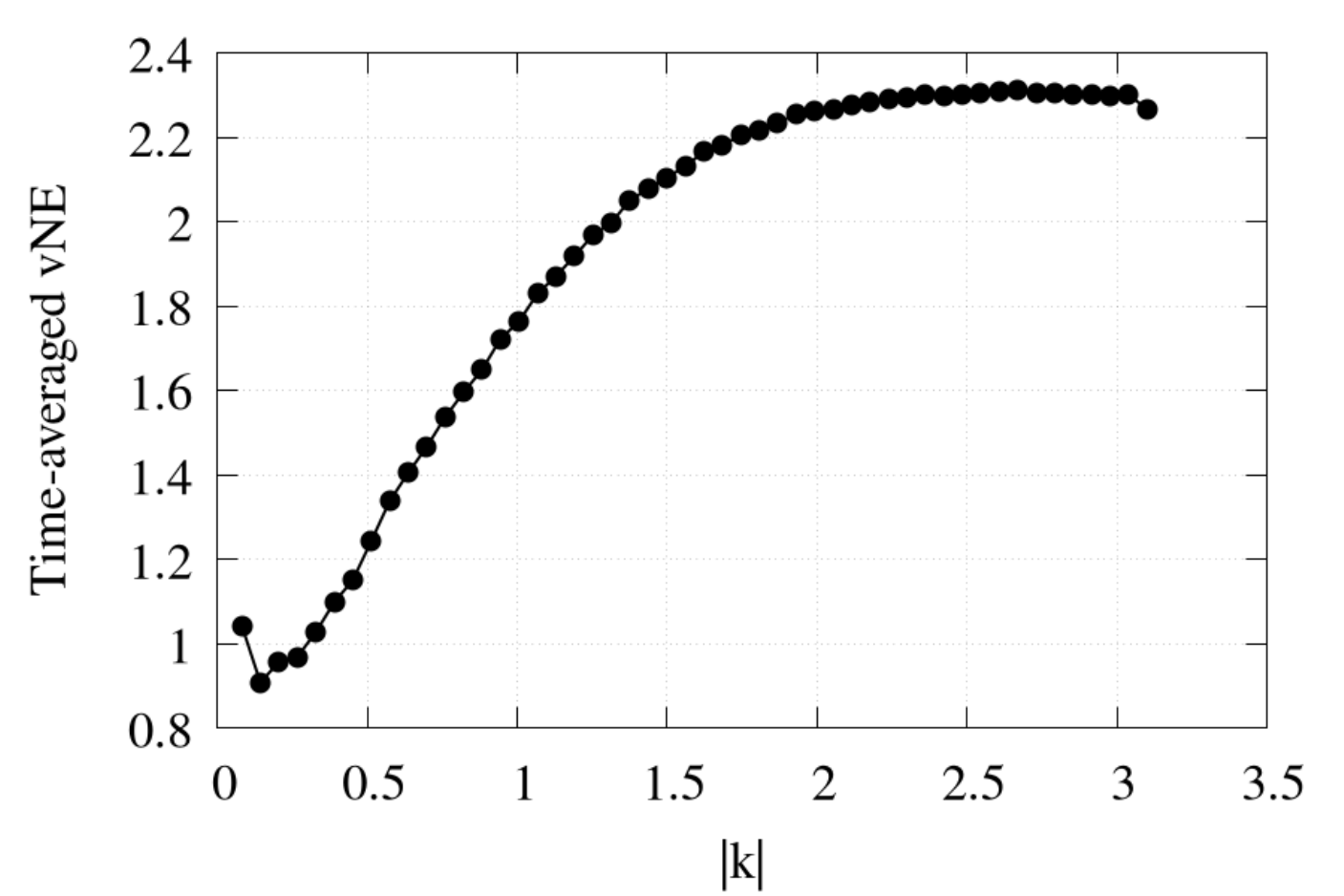}
  \caption{The dependence of the von Neumann entropy (vNE) of the gyrokinetic distribution function on the absolute value of the wavenumber $|\vb*{k}|$.} 
  \label{fig:kabs-vNE}
\end{figure}

Finally, Fig.~\ref{fig:kabs-vNE} plots the vNE against $|\vb*{k}|$. The curve exhibits a rapid increase across $|\vb*{k}|\!\sim\!1$, reinforcing the view that the saturated turbulence reorganizes from a low-rank, coherent regime at large scales to a high-rank, strongly mixed regime at smaller scales. \par
We reveal a nontrivial, wavenumber dependence in velocity-space complexity characterized by the vNE. Because vNE measures how many data-driven modes are needed to represent the field (without presupposing a Fourier or PDF basis), this provides a compact, basis-agnostic indicator that uncovers previously overlooked structure in gyrokinetic turbulence.

\section{Discussion} \label{sec:discussion}
In this section, we interpret the wavenumber dependence of velocity-space complexity revealed by the vNE analysis by confronting it with linear physics and reduced velocity-space diagnostics. \par
\begin{figure}[H]
  \centering
  \includegraphics[width=12cm]{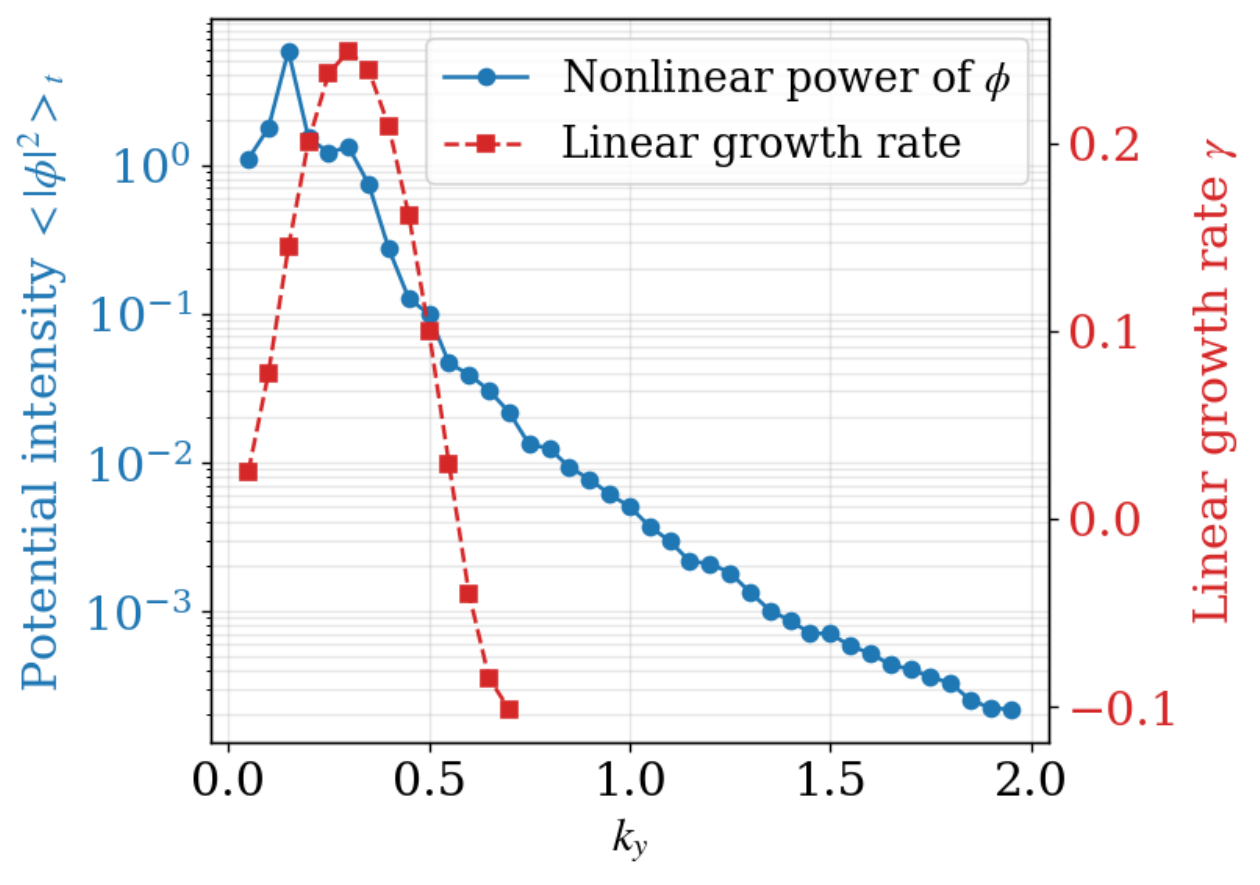}
  \caption{The $k_y$ spectrum of the electrostatic potential intensity $\langle|\phi|^2\rangle_t$ in the nonlinear simulation of GKV (blue and left axis) and the growth rate $\gamma_{\mathrm{lin}}$ in the linear simulation of GKV (red and right axis).} 
  \label{fig:ky-gamma-phi}
\end{figure}

Figure~\ref{fig:ky-gamma-phi} compares the linear gyrokinetic growth rate $\gamma_{\mathrm{lin}}(k_y)$ from a linear simulation of GKV with the nonlinear saturated potential power at $k_x\!=\!0$ as a function of $k_y$. While a slight spectral shift to the low-$k$ region from the peak in the linear spectrum, the nonlinear power peak still remains around the band of large $\gamma_{\mathrm{lin}}$. This indicates that the low-$k$ portion of the saturated turbulence is mainly supplied by ITG-unstable modes, consistent with established ITG phenomenology \cite{Dimits2000}. This ITG dominance at low $k$ naturally explains the smaller vNE there: a few dominant SVD modes capture most of the variance, while higher-$k$ ranges require many modes due to enhanced phase-space mixing. \par

To resolve phase-space structure of the distribution function, we project the real component of the perturbed distribution function
\(\Re[\delta f_{\vb*{k}}(\xi,\zeta,t)]\) onto orthonormal bases along
\(\xi\equiv v_\parallel/v_{\mathrm{th}}\in\mathbb{R}\) and
\(\zeta\equiv \mu/\mu_{\mathrm{th}}\in[0,\infty)\).
The normalized bases are built from the Hermite polynomial and the Laguerre polynomial defined as
\begin{align}
  H_m(\xi) &= \frac{(-1)^{m}}{(2\pi)^{1/4}\sqrt{m!}} e^{\xi^{2}/2}\,\dv[m]{\xi}\!\left(e^{-\xi^{2}/2}\right), 
  & \int_{-\infty}^{\infty}\!\dd{\xi}\,e^{-\xi^{2}/2}\,H_{m}(\xi)\,H_{n}(\xi)=\delta_{mn},\\
  L_\ell(\zeta) &= e^{\zeta}\,\frac{1}{\ell!}\,\dv[\ell]{\zeta}\!\left(\zeta^{\ell}e^{-\zeta}\right),
  & \int_{0}^{\infty}\!\dd{\zeta}\,e^{-\zeta}\,L_{k}(\zeta)\,L_{\ell}(\zeta)=\delta_{k\ell}.
\end{align}
With this choice the Hermite set is explicitly normalized under the Gaussian weight, while the Laguerre set is the standard orthonormal system under the exponential weight.
At fixed \(\zeta\) and \(\xi\), the Hermite and the Laguerre coefficients are calculated as
\begin{align}
  F_{\mathrm{Her}}^{m}(\vb*{k};\zeta,t)
    &= \int_{-\infty}^{\infty}\!\dd{\xi}\,e^{-\xi^{2}/2}\,
       H_{m}(\xi)\,\delta f_{\vb*{k}}(\xi,\zeta,t), \\
  F_{\mathrm{Lag}}^{\ell}(\vb*{k};\xi,t)
    &= \int_{0}^{\infty}\!\dd{\zeta}\,e^{-\zeta}\,
       L_{\ell}(\zeta)\,\delta f_{\vb*{k}}(\xi,\zeta,t).
\end{align}
The corresponding modal intensities (including the complementary integration in the other velocity coordinate and a time average \(\langle\cdot\rangle_t\)) are
\begin{align}
  W_{\mathrm{Her}}^{m}(\vb*{k})
    &= \Big\langle \int_{0}^{\infty}\!\dd{\zeta}\,e^{-\zeta}\,
       \abs{F_{\mathrm{Her}}^{m}(\vb*{k};\zeta,t)}^{2}\Big\rangle_{t}, \\
  W_{\mathrm{Lag}}^{\ell}(\vb*{k})
    &= \Big\langle \int_{-\infty}^{\infty}\!\dd{\xi}\,e^{-\xi^{2}/2}\,
       \abs{F_{\mathrm{Lag}}^{\ell}(\vb*{k};\xi,t)}^{2}\Big\rangle_{t}.
\end{align}
\noindent
These definitions provide rigorously orthonormal, Maxwellian-weighted spectral diagnostics: growth of \(W_{\mathrm{Her}}^{m}\) toward large \(m\) quantifies parallel phase mixing (Landau resonances), while growth of \(W_{\mathrm{Lag}}^{\ell}\) toward large \(\ell\) captures perpendicular FLR phase mixing. \par
\begin{figure}[H]
  \centering
  \includegraphics[width=16cm]{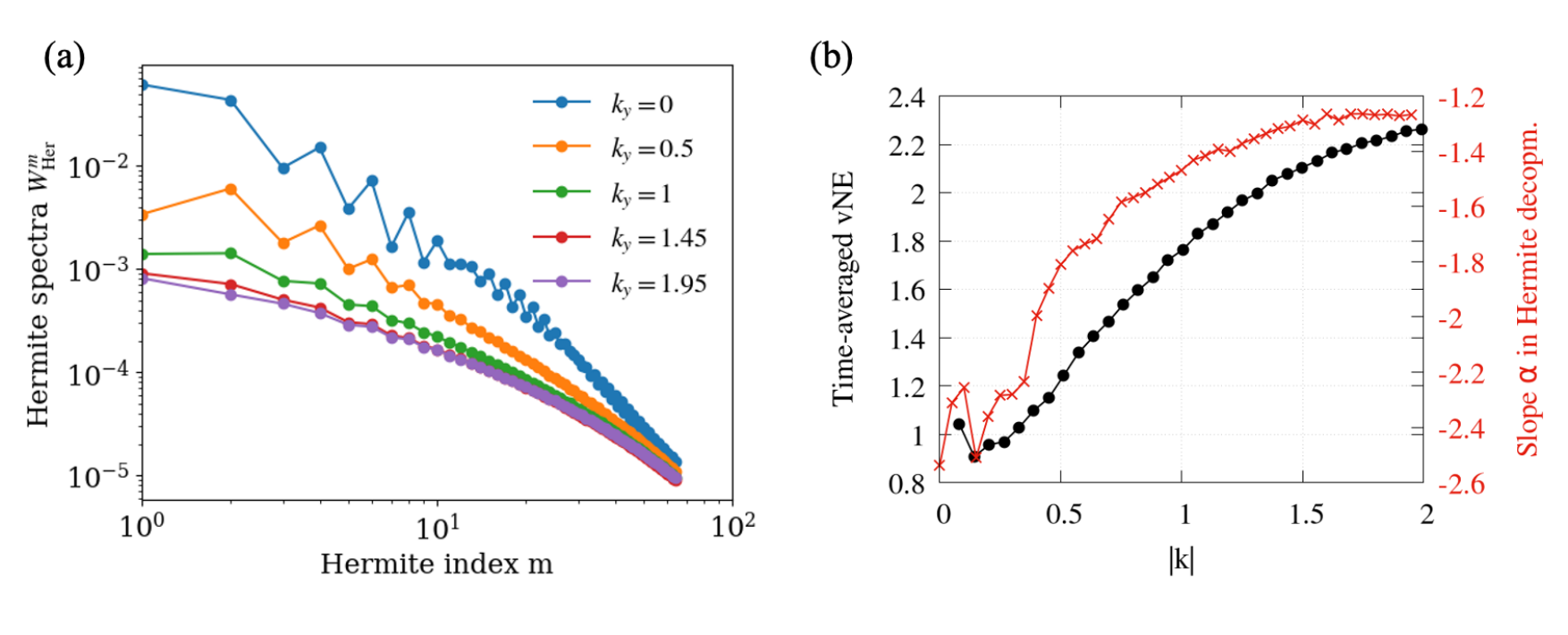}
  \caption{(a) The Hermite spectrum of the modal intensities $W_{\mathrm{Her}}^m$ of the Hermite decomposition of the distribution function for several values of $k_y=0,0.5,1,1.45,1.95$. (b) The $k_y$ dependence of the slope $\alpha$ in the large $m$ spectrum of $W_{\mathrm{Her}}^m$.} 
  \label{fig:ky-WHer}
\end{figure}

Figure~\ref{fig:ky-WHer}(a) shows the Hermite spectra $W_{\mathrm{Her}}^{m}(\vb*{k})$ at $k_x=0$ for several values of $k_y$. At large $m$ the spectra decay approximately as a power law, $W_{\mathrm{Her}}^{m}\propto m^{-\alpha}$, while the tail becomes progressively shallower with increasing $k_y$. This trend indicates that parallel phase mixing (Landau resonances) becomes more active at higher perpendicular wavenumber: energy is transferred toward finer structure in $v_\parallel$-space, populating higher Hermite orders. Consistently, the vNE rises with $|\vb*{k}|$ because more SVD modes are required to represent the data. Quantitatively, the $k_y$-dependence of the tail slope $\alpha$, plotted in Fig.~\ref{fig:ky-WHer}(b) from fits to the large-$m$ range, closely follows the $k_y$-dependence of the time-averaged vNE. This correspondence supports the interpretation that the observed increase of vNE across wavenumber space is predominantly caused by enhanced parallel phase mixing that drives a broader Hermite spectrum. \par
\begin{figure}[H]
  \centering
  \includegraphics[width=12cm]{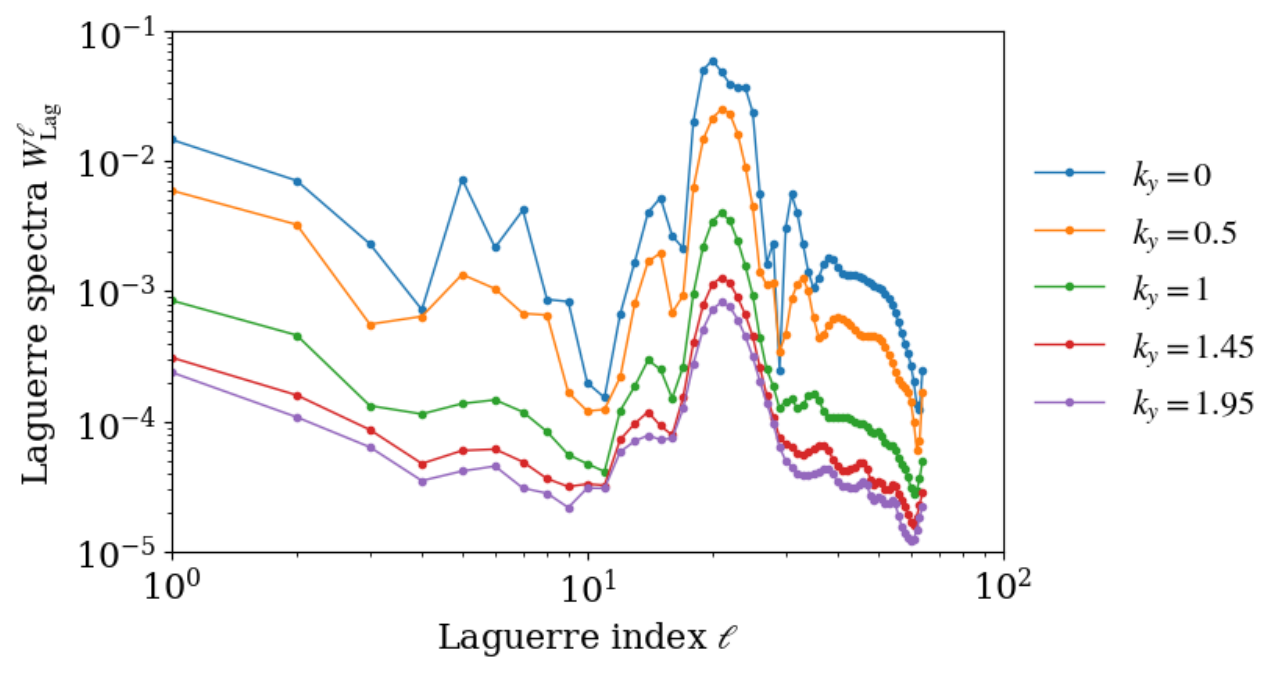}
  \caption{The Laguerre spectrum of the modal intensities $W_{\mathrm{Lag}}^\ell$ of the Laguerre decomposition of the distribution function for several values of $k_y=0,0.5,1,1.45,1.95$.} 
  \label{fig:ky-WLag}
\end{figure}

Figure~\ref{fig:ky-WLag} presents the Laguerre spectra $W_{\mathrm{Lag}}^{\ell}(\vb*{k})$ at $k_x=0$ for several $k_y$. All curves exhibit a pronounced peak around $\ell\sim20$, indicating active perpendicular FLR phase mixing that builds structure in the $\mu$ direction. In contrast to the Hermite case, the overall shape of the Laguerre spectra (e.g., the ratio of low-$\ell$ to high-$\ell$ content and the tail steepness) varies only weakly with $k_y$ within the present parameter range. Then, it is indicated that the perpendicular FLR phase mixing has a limited impact on the strong wavenumber dependence seen in the vNE maps (Figs.~\ref{fig:kx-ky-vNE} and \ref{fig:kabs-vNE}) in this formalism. \par
The results in this study do not fully agree with the standard theoretical picture: the nonlinear FLR phase mixing becomes more pronounced for $|\vb*{k}| > 1$ in comparison with the linear Landau phase mixing. On the other hand, Parker \textit{et al.} demonstrated the effect of ''anti-phase-mixing'', which suppresses the linear Landau resonances due to the nonlinear advection, in the ITG-driven drift-kinetic equation \cite{Parker2016}. Especially, Fig.~3 in Ref.~\cite{Parker2016} shows the steep slope at the high perpendicular wavenumber case compared to the low perpendicular wavenumber case. However, according to Ref.~\cite{Schekochihin2016}, the balance between the Landau resonances and the anti-phase-mixing is determined by the ratio between the linear streaming $k_\parallel v_\mathrm{th}$ and the nonlinear decorrelation rate $k_\perp u_\perp$, i.e., the balance depends on not only the wavenumber $(k_\parallel,k_\perp)$ but also the $\vb*{E}\times\vb*{B}$ velocity $u_\perp$, which is proportional to the intensity of the turbulence. In the present study, the amplitude of the electrostatic potential at high perpendicular wavenumber is about two orders of magnitude smaller than at low perpendicular wavenumber, as shown in Fig.~\ref{fig:t-E}. Thus, there remains the possibility that Landau resonances may overcome anti-phase-mixing due to small $\vb*{E}\times\vb*{B}$ velocity, even with a high perpendicular wavenumber.


\section{Summary and Outlook} \label{sec:summary}
In this paper, we have developed and applied a diagnostics framework that combines singular value decomposition (SVD) with an information-theoretic weighting to quantify the velocity-space complexity of gyrokinetic turbulence, which is inspired from the analogy between the SVD and the quantum states. Using nonlinear flux-tube simulations of the gyrokinetic-Poisson equation, we computed the von~Neumann entropy (vNE) of the time-evolving ion distribution function at each wavenumber $(k_x,k_y)$ and found a rapid increase of vNE across $|\vb*{k}|\sim 1$: low-wavenumber modes are representable by a few SVD modes (small vNE), whereas high-wavenumber modes require many modes (large vNE) to restore the original structure of the distribution function. Comparison with linear growth rates shows that the low-wavenumber peak of the nonlinear electrostatic potential power remains in the vicinity of the ITG unstable range. Hermite-Laguerre diagnostics suggest that the vNE increase at high wavenumber is primarily associated with enhanced parallel phase mixing i.e., Hermite-order transfer, with a comparatively weaker dependence on perpendicular FLR mixing, whereas the detailed comparison with theoretical predictions from prior research is further required. These results expose a previously uncharted, scale-dependent variation of phase-space complexity across wavenumbers and establish vNE as a compact, data-driven measure of how many dynamically active degrees of freedom are needed to represent kinetic turbulence. \par
The present vNE-based methodology opens a path to extract concrete phase-space mechanisms and their influence on the transport of particles and energy. A possible next step is to target phase-space coherent structures such as BGK vortices~\cite{Bernstein1957,Lesur2014} within nonlinear gyrokinetic dynamics: by combining vNE maps with Fourier-Hermite/Laguerre spectra and cross-phase/coherence analyses, we can verify whether intermittently formed phase-space islands act as reservoirs or conduits of free energy and how they bias cross-field transport. Systematic scans in collisionality, drive, interplay of parallel and perpendicular advection for the high-$k$ region, and geometry (including electron-scale and multi-species effects) will clarify when phase mixing is fluidized or restored, and how those changes imprint on vNE and on experimentally relevant transport channels.

\section*{Acknowledgements}
Numerical computations are performed on the NIFS Plasma Simulator. This work was supported by the NIFS collaborative Research Programs (NIFS23KIST039, NIFS23KIST044, NIFS25KISM011, NIFS24KIPM005).


\section*{References}
\bibliography{vNE_ITG}

\end{document}